\documentclass[letterpaper]{article} 
\usepackage{aaai24}  
\usepackage{times}  
\usepackage{helvet}  
\usepackage{courier}  
\usepackage[hyphens]{url}  
\usepackage{graphicx} 
\usepackage{natbib}  
\usepackage{caption} 
\usepackage{algorithm}
\usepackage{algorithmic}

\usepackage{newfloat}
\usepackage{listings}
\DeclareCaptionStyle{ruled}{labelfont=normalfont,labelsep=colon,strut=off} 
\floatstyle{ruled}
\newfloat{listing}{tb}{lst}{}
\floatname{listing}{Listing}
\usepackage{amssymb}
\usepackage{booktabs}

\title{Pre-training with Large Language Model-based Document Expansion \\for Dense Passage Retrieval}
\author{
    Guangyuan Ma\textsuperscript{\rm 1,2}\equalcontrib,
    Xing Wu\textsuperscript{\rm 1,2}\equalcontrib,
    Peng Wang\textsuperscript{\rm 1,2},
    Zijia Lin\textsuperscript{\rm 3},
    Songlin Hu\textsuperscript{\rm 1,2}
}
\affiliations{

    \textsuperscript{\rm 1} Institute of Information Engineering, Chinese Academy of Sciences, Beijing, China \\
    \textsuperscript{\rm 2} School of Cyber Security, University of Chinese Academy of Sciences, Beijing, China \\
    \textsuperscript{\rm 3} Kuaishou Technology \\
    \{maguangyuan,wuxing,wangpeng2022,husonglin\}@iie.ac.cn, linzijia07@tsinghua.org.cn
}

\begin{document}

\maketitle

\begin{abstract}

In this paper, we systematically study the potential of pre-training with Large Language Model(LLM)-based document expansion for dense passage retrieval. Concretely, we leverage the capabilities of LLMs for document expansion, i.e. query generation, and effectively transfer expanded knowledge to retrievers using pre-training strategies tailored for passage retrieval. These strategies include contrastive learning and bottlenecked query generation. Furthermore, we incorporate a curriculum learning strategy to reduce the reliance on LLM inferences. Experimental results demonstrate that pre-training with LLM-based document expansion significantly boosts the retrieval performance on large-scale web-search tasks. Our work shows strong zero-shot and out-of-domain retrieval abilities, making it more widely applicable for retrieval when initializing with no human-labeled data. 
\end{abstract}

\section{Introduction}
Dense passage retrieval \cite{karpukhin2020dense} has broad real-world applications, like web search \cite{Liu2021baidu_search, Zou2023pre_trained}, retrieval-augmented generation \cite{Lewis2020Retrieval, cai2022recent} and query answering \cite{Sakata2019FAQ}. It utilizes well-trained language-model-based retrievers to extract sentence representations and retrieve relevant passages with given queries.
Recent studies have made impressive progress in improving the effectiveness of dense retrievers, such as hard negative mining \cite{qu-etal-2021-rocketqa}, late interaction \cite{Omar2020colbert, santhanam2022colbertv2}, distillation \cite{ren-etal-2021-rocketqav2, lu2022ernie_search}, and ensembling \cite{gao2022unsupervised, wu2023cotmaev2}. Moreover, the development of task-specific pre-training \cite{gao2021condenser, wu2023contextual, liu2022retromae} pushes the limits of retrieval tasks to new boundaries. Specifically, those studies usually employ contrastive learning with span corruption \cite{gao2022unsupervised, Izacard2021contriever, ma2022pre}, or additional decoder with bottlenecked structures \cite{gao2021condenser, lu2021less, liu2022retromae, wu2023contextual} for better representation learning.

Large language models (LLMs), like ChatGPT \cite{long2022instructgpt}, PaLM \cite{Aakanksha2022PaLM}, LLaMA \cite{Hugo2023LLaMA}, and tk-Instruct \cite{Yizhong2022SuperNaturalInstructions}, are pre-trained on large-scale web corpus and exhibit excellent abilities in context generation and instruction following. 
There has been growing interest in incorporating powerful LLMs into retrieval tasks. Existing studies \cite{gao2023hyde, wang2023Query2doc, jagerman2023QueryExpension, yu2023Generate} focus on \textit{query expansion} with LLMs for enhancing the lexical match of query-passage pairs. They utilize the LLM-generated relevant passages as enriched query contexts. Those studies have yielded better retrieval performances, especially for zero-shot scenarios. Nevertheless, conducting query expansion still needs heavy online inferences with LLMs, which slows down the retrieval speed.

While query expansion expands the query with generated passages, \textit{document expansion}, i.e., query generation, is also a popular technique to boost retrieval performances. It exploits a fully fine-tuned model, like T5 \cite{Rodrigo2019doc2query} or BART \cite{Cho2022QueryGeneration}, to generate relevant queries of a given passage, which either enrich the context of the passage or serve as additional fine-tuning corpus. Due to the excellent generation ability of LLMs, huge potential lies in the utilization of LLMs as document expansion models. However, we argue that several drawbacks still hinder such usage. 
Firstly, document expansion relies on the online inference of LLM in open-domain passage retrieval, particularly when dealing with candidate corpora from new domains. To avoid the need for additional LLM inferences during retrieval, a feasible solution is to pre-train or fine-tune an end-to-end retriever. However, this approach lacks exploration and necessitates training paradigms specifically designed for retrieval.
Furthermore, document expansion involves feeding a substantial corpus into LLMs to generate queries, resulting in significant costs associated with LLM inferences. Unfortunately, there is a shortage of methods to mitigate these inference costs.

\begin{figure}[t!]
\centering
\includegraphics[width=\linewidth]{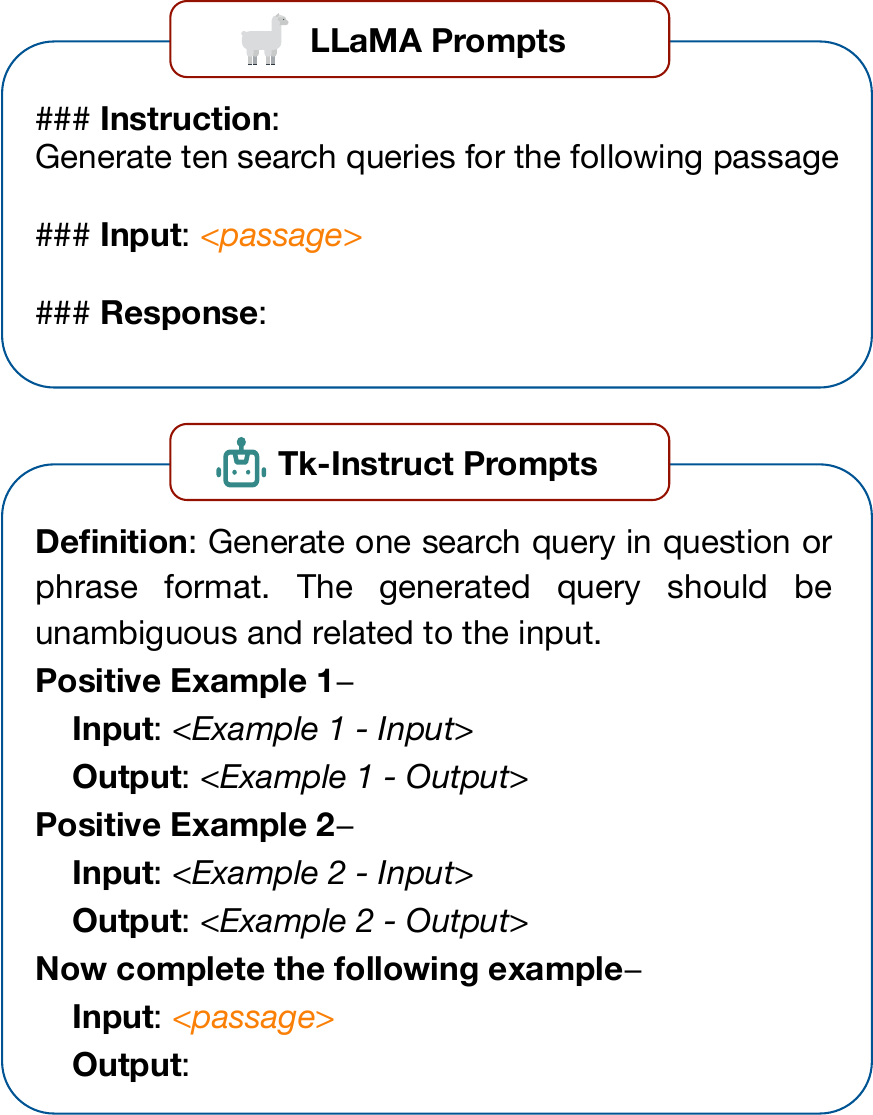}
\caption{
Query Generation prompts for Alpaca-LLaMA and tk-Instruct.
}
\label{llm_prompts}
\end{figure}

To mitigate the high online inference costs of LLM document expansion, as is presented in Figure \ref{llm_prompts}, we prompt the LLM query generation for a series of pre-training experiments tailored for dense retrieval. We emphasize that our work only involves LLM inferences at the pre-training stage of retrievers, but not the inference stage as traditional query \cite{gao2023hyde, wang2023Query2doc} or document expansion \cite{Rodrigo2019doc2query}. Two pre-training paradigms, i.e., contrastive learning and bottlenecked query generation, are explored in detail.

For contrastive pre-training, a direct contrastive loss of the generated queries and passages is used to pull together their embeddings, while pushing away in-batch negatives in the latent space. We follow the contrastive architecture in  \cite{gao2022unsupervised} for fair comparision, and we argue that LLM-generated queries can serve as the better context for effective query-passage alignment.

Bottlenecked pre-training techniques are popular in recent works \cite{lu2021less, liu2022retromae, wu2023contextual}, which connect accessional decoders solely through the encoder's representation. To pre-train with bottlenecked query generation, similar to \cite{wu2022query-as-context}, we adapt a single-layer Transformers decoder and use the casual language model (CLM) task to generate expanded queries with the assistance of the encoder's embeddings. This bottlenecked encoder-decoder structure first derives a compressed representation through the encoder and then decompresses the context information as LLM-expanded queries via the decoder. As a result, the sentence embeddings contain enriched context information, providing effective initialization for fine-tuning and inference. Especially, LLM-based document expansion requires no human-labeled corpus as previous works \cite{wu2022query-as-context, Cho2022QueryGeneration} for training additional domain-specific generative models like docT5query \cite{Rodrigo2019doc2query}. 

Furthermore, to mitigate the LLM inference costs for document expansion, we interpolate a two-stage curriculum learning strategy for both pre-training schemas. Span corruption is firstly used to randomly sample contextual pairs from a long document. Then we leverage the generation abilities of LLMs to produce a relatively small amount of queries for the next stage of pre-training. 

In our study, we use Alpaca-LLaMA \cite{wang2023selfInst} and tk-Instruct \cite{Yizhong2022SuperNaturalInstructions} with different parameter sizes for query generation. We conduct the experiments on the large-scale MS-MARCO \cite{tri2016msmarco} datasets and test on the in-domain MS-MARCO passage retrieval task, TREC-DL 2019 \& 2020 \cite{craswell2020trec19, craswell2021trec20} and the out-of-domain BEIR \cite{thakur2021beir} task. 
Several benefits are observed in our studies. 1) LLMs can generate a large number of high-quality queries based on the world knowledge of LLM itself, which requires no additional human labeling and is suitable for scenarios lacking in manually annotated data. 2) Contrastive pre-training with LLM-generated queries has stronger in-domain zero-shot retrieval performance and on-par performance with the state-of-the-art (SOTA) methods after full fine-tuning. It also shows better domain adaption abilities in out-of-domain BEIR datasets. 3) Bottlenecked query generation shows better initialization abilities after full fine-tuning. 4) With our two-stage curriculum learning strategy, we reduce the number of MS-MARCO corpus involved in LLM inferences from 8.8 million to 0.4 million, while keeping the minor performance degeneration. 

Our contributions are summarized as follows.
\begin{enumerate}
\item[•] We systematically study the potential of incorporating LLMs into the pre-training stage of dense passage retrieval, suitable for the scarcity of human-annotated data.
\item[•] We find stronger zero-shot and fine-tuned performances with contrastive learning and good initialization abilities with bottlenecked query generation pre-training.
\item[•] We design a two-stage curriculum learning strategy that greatly reduces the usage of LLM-expanded queries while keeping the minor performance degeneration.
\end{enumerate}

\begin{figure*}[t!]
\centering
\includegraphics[width=\linewidth]{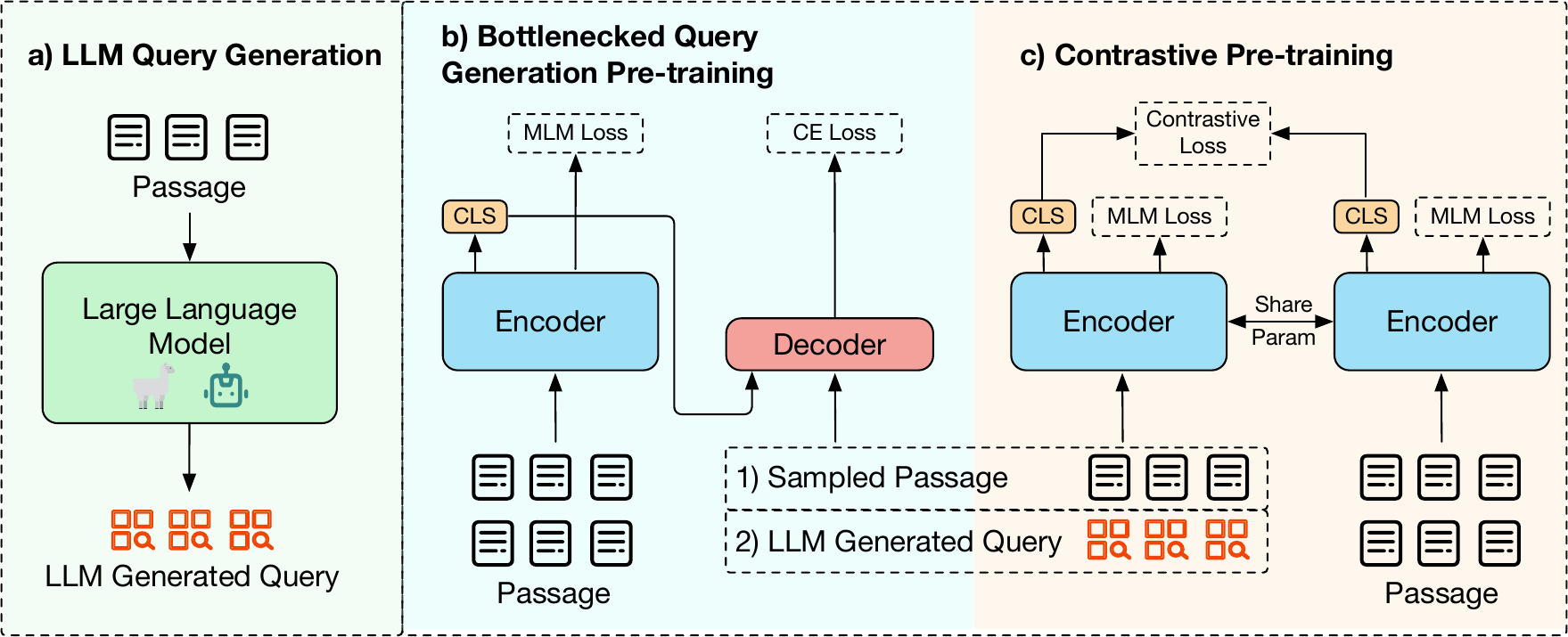}
\caption{
Pre-training with LLM-based document expansion for dense passage retrieval. \textbf{a)} We utilize large language models (LLMs) to generate pseudo-queries with zero-shot or few-shot prompts. \textbf{b)} Bottlenecked query generation pre-training appends an auxiliary Transformers decoder to the encoder. Besides the Masked Language Modelling (MLM) loss of the encoder, we connect the encoder-decoder with merely the bottlenecked representation, i.e., the hidden states of [CLS] token, and make the decoder generate whole LLM-expanded queries with the Cross-Entropy (CE) loss. \textbf{c)} Contrastive pre-training pulls together the representations of the passage and LLM-expanded queries and pushes away in-batch negatives. To minimize reliance on LLM expansions, we implement a two-stage curriculum learning strategy. It first utilizes randomly sampled passages to fully initialize the encoders. And then we can use a relatively small amount of LLM-expanded queries in the second phase. 
}
\label{model}
\end{figure*}

\section{Methodology}
In this section, we first introduce the definition of dense passage retrieval. Then we introduce our method for LLM query generation, the detailed pre-training designs of contrastive learning and bottlenecked query generation, and the two-stage curriculum learning strategy for extended analyses.

\subsection{Preliminaries}
Given a query $q$ and a set of passages $P_n$, the passage retrieval task aims to find the relevant passages based on the similarity search. Dense passage retrieval utilizes an encoder model $Enc$, e.g., a Transformers-based model like BERT \cite{devlin2019bert}, to yield the sentence representations and measure query-passage similarities through inner product or cosine distance. Formally, given a query $q$ and a passage $q$, we can use a query encoder $Enc_q$ and a passage encoder $Enc_p$ to derive their corresponding sentence representations, i.e., $v_q$ and $v_p$ from the encoder hidden states of the last layer at CLS position $h_{last}^{\texttt{[CLS]}}$. Then the similarity between $q$ and $p$, i.e., $Sim(q, p)$, can be calculated as the inner product of $v_q$ and $v_p$ for simplicity as follows.

\begin{equation}
\label{equation_query_passage_similarity}
    Sim(q, p) = Enc_{q}(q) \cdot Enc_{p}(p) = v_q^T v_p
\end{equation}

The key to improving retrieval performances is to yield stronger representations $v_q$, $v_p$ with better context alignment. The representations can be regarded as the compression of full contexts. We believe that incorporating the strong context-generation abilities of LLMs into the pre-training stage with carefully designed pre-tasks can be a new way for improving such alignment.

\subsection{LLM Query Generation}
Given a passage $p$, we use a zero-shot prompt for Alpaca-LLaMA and a few-shot prompt for tk-Instruct to expand queries, as illustrated in Figure \ref{llm_prompts}. We empirically find that Alpaca 7B and 13B models work well on the zero-shot prompt, which helps save computation budgets. We manually write a few examples for tk-Instruct, as we find that few-shot prompts make its query generation more stable.

LLM-based document expansion enriches the pre-training corpus with additional contextual information. Instead of directly appending the expanded queries onto the passage, we seek to incorporate them into our pre-training stage for better initialization of end-to-end retrievers. Our work only involves LLM inference at the pre-training stage, but not the retrieval stage like traditional query or document expansion works.
Two pre-training paradigms are involved to incorporate the LLM-generated queries into the dense model pre-training.

\begin{table*}[!htbp]
\centering
\begin{tabular}{l|ccc|c|c}
\toprule
 ~ & \multicolumn{3}{c|}{\textbf{MS-MARCO}} & \textbf{TREC DL 19} & \textbf{TREC DL 20}\\
\textbf{Model / Zero-shot Evaluation} & MRR@10 & Recall@50 & Recall@1k & nDCG@10 & nDCG@10 \\
\midrule
BM25 &  18.7  & 59.2  & 85.7  & 51.2 & 47.7 \\
SimCSE \cite{gao2021simcse}$\dagger$ & 8.7 & 33.7 & 64.6 & 24.5 & 17.9 \\
coCondenser \cite{gao2022unsupervised}$\dagger$ & 7.5 & 31.3 & 58.1 & 22.1 & 20.7 \\
Contriever \cite{Izacard2021contriever}$\dagger$ & 16.8 & 60.8 & 89.1 & 44.5 & 43.2 \\
\midrule
\multicolumn{6}{l}{\textbf{\textit{Contrastive Pre-training}}} \\ 
\midrule
Baseline & 12.5 & 49.0 & 82.3 & 36.0 & 38.4 \\
 + tk-inst 3b queries & 20.9\textsuperscript{+8.4} & 70.2\textsuperscript{+21.2} & 92.8\textsuperscript{+10.5} & 47.0\textsuperscript{+11.0} & 48.6\textsuperscript{+10.2} \\
 + Alpaca 7b queries & 22.6\textsuperscript{+10.1} & 70.7\textsuperscript{+21.7} & 93.8\textsuperscript{+11.5} & 51.0\textsuperscript{+15.0} & 48.9\textsuperscript{+10.5} \\
 + Alpaca 13b queries & \textbf{22.7\textsuperscript{+10.2}} & \textbf{71.7\textsuperscript{+22.7}} & \textbf{94.3\textsuperscript{+12.0}} & \textbf{53.9\textsuperscript{+17.9}} & \textbf{50.1\textsuperscript{+11.7}} \\
\bottomrule
\end{tabular}
\caption{
Zero-shot evaluation of contrastive pre-training with LLM-based document expansion.
$\dagger$ denotes our reproduced results. 
The best scores are marked in \textbf{bold}. Results with the increment over the corresponding baseline have been tested with two-tailed t-tests, demonstrating statistically significant improvements ( p-value $\leq$ 0.01 ).
}
\label{table_results_zero_shot}
\end{table*}

\subsection{Bottlenecked Query Generation Pre-training}
Bottlenecked pre-training trains a monomeric encoder ($Enc$) with good initialization abilities for subsequent fine-tuning. Given a tokenized sentence $t \in T$ from the training corpus, we randomly select a certain ratio of tokens, with the corresponding indices denoted as $M$, and replace them with mask tokens $\texttt{[m]}$:
\begin{equation}
    mask(t) = \{\texttt{[CLS]}, t_1, t_2, \texttt{[m]}, t_4, ..., t_n, \texttt{[SEP]}\}
\end{equation}

Cross-Entropy (CE) loss is then used to optimize as Masked Language Model (MLM) loss for the encoder.

\begin{equation}
\mathcal{L}_{enc}=-\sum_{t\in T}\sum_{i \in {M}} \log p(t_i|Enc(mask({t})))
\end{equation}
where $t_i$ is groundtruth tokens w.r.t corresponding mask tokens $\texttt{[m]}$.

A single-layer accessional Transformers decoder ($Dec$) is further introduced, which receives the input from the concatenation of the encoder representation $h_{last}^{\texttt{[CLS]}}$ and contextual texts $x$, e.g., LLM-generated queries.

\begin{equation}
{T}_{ctx} = \{h_{last}^{\texttt{[CLS]}}, x_1, ..., x_N, \texttt{[SEP]}\}
\end{equation}

Then the decoder uses the Casual Language Model (CLM) loss to generate the whole input context with the assistance of encoder representation.

\begin{equation}
\mathcal{L}_{dec}=-\sum_{x_{i} \in {T}_{ctx}} \log p(x_{i}|Dec(x[:i-1]))
\end{equation}

The final loss $\mathcal{L}$ is then formulated as follows.

\begin{equation}
\mathcal{L}=\mathcal{L}_{enc} + \mathcal{L}_{dec}
\end{equation}

Through the bottlenecked encoder-decoder structure, we seek to compress the context signal from LLM-generated queries into the encoder representations and give strong initialization ability to the encoder.

\subsection{Contrastive Pre-training}
For reproduction and fair comparison, we adapt the contrastive pre-training architecture from coCondenser \cite{gao2022unsupervised}. The passage $p$ and its sampled or generated context ${p}_{ctx}$ are directly forwarded through the encoder $Enc$. Besides the MLM loss $\mathcal{L}_{enc}$ of the encoder, an extra Transformers decoder $Dec_{ext}$ is also introduced for representation pre-training, which takes the concatenation of the encoder representation $h_{last}^{\texttt{[CLS]}}$ and encoder hidden states $h_{l}^{i}$ from $l$-th layer. Then a cross-entropy loss is used for the decoder's pre-task.

\begin{equation}
\mathcal{L}_{ext}=-\sum_{t\in T}\sum_{i \in {M}} \log p(t_i|Dec_{ext}({h_{last}^{\texttt{[CLS]}}, h_{l}^{1}, ..., h_{l}^{i}}))
\end{equation}

Differently, for pre-training with LLM-expanded queries, assuming $v_p$ and $v_{ctx}$ denote encoders' representations, a contrastive loss with in-batch negatives is used as follows.

\begin{equation}
\mathcal{L}_{CL}=-\log \frac{\exp(v_p \cdot v_{ctx}^{+})}{\exp({v_p \cdot v_{ctx}^{+}}) + \sum \exp({v_p \cdot v_{ctx}^{-}})}
\end{equation}

\noindent where $v_{ctx}^{+}$ is the context text, e.g. LLM-generated queries, corresponding to $p$. And $v_{ctx}^{-}$ is the in-batch negatives, i.e., the context texts of the other passages in the batch.

The final optimization objective is the sum of the above losses.

\begin{equation}
\mathcal{L}=\mathcal{L}_{enc} + \mathcal{L}_{ext} + \mathcal{L}_{CL}
\end{equation}

Through contrastive pre-training, the representations of passage and LLM-generated queries are directly pulled together in the same latent space, which gives better query-passage alignment and zero-shot ability to encoders.

\subsection{Curriculum Learning}
As discussed before, LLM-based document expansion faces the challenge of costly inference due to large numbers of documents or passages. 
Since we intend to pre-train our model with enriched contexts, inspired by the wisdom of curriculum learning \cite{Bengio2009curriculumLearning}, we consider 1) a randomly cropped passage span as a coarse-grained context, while 2) the LLM-expanded queries as fine-grained context, as depicted in Figure \ref{model}. 
Following the span corruption strategies in the seed-encoder \cite{lu2021less} and coCondenser \cite{gao2022unsupervised}, we use the coarse-grained context as the passage itself in the bottlenecked generation pre-training, and the randomly sampled passage span in contrastive pre-training. As we focus on LLM-based document expansion, other span corruption strategies \cite{wu2023contextual} are left to our future work.
After pre-training on a large amount of randomly cropped contexts, we initialize from the first stage and then use the fine-grained LLM-expanded queries for the second-phrase pre-training. Experiments find that this curriculum strategy greatly reduces the need for LLM inferences on MS-MARCO passages, while still maintaining similar retrieval performances.

\begin{table*}[!htbp]
\centering
\begin{tabular}{l|ccc|c|c}
\toprule
 ~ & \multicolumn{3}{c|}{\textbf{MS-MARCO}} & \textbf{TREC DL 19} & \textbf{TREC DL 20}\\
\textbf{Model / Fine-tuned Results} & MRR@10 & Recall@50 & Recall@1k & nDCG@10 & nDCG@10 \\
\midrule
Contriever \cite{Izacard2021contriever}$\dagger$ & 33.4 & 85.0 & 98.4 & 62.8 & 63.2 \\
Condenser \cite{gao2021condenser} & 36.6  & 85.4$\dagger$ & 97.4  & 69.8 & 66.5$\dagger$ \\
coCondenser \cite{gao2022unsupervised} & 38.2  & 86.5$\dagger$  & 98.4  & 71.7$\dagger$ & 68.4$\dagger$ \\
SimLM \cite{wang2022simlm} & 39.1 & 87.3$\dagger$ & 98.6  & 68.9$\dagger$ & 68.8$\dagger$ \\
RetroMAE \cite{liu2022retromae} & 39.3  & 87.0$\dagger$ & 98.5  & 69.1$\dagger$ & 70.0$\dagger$ \\
CoT-MAE \cite{wu2023contextual} & 39.4 & 87.0 & 98.7 & 70.9$\dagger$ & 70.4 \\
\midrule
\multicolumn{6}{l}{\textbf{\textit{Contrastive Pre-training}}} \\ 
\midrule
Baseline & 38.8  & 87.8 & 98.8 & 71.1 & 68.4 \\
 + tk-instruct 3b queries & 39.6\textsuperscript{+0.8} & 88.8\textsuperscript{+1.0} & 99.0 & \textbf{72.9\textsuperscript{+1.8}} & 71.1\textsuperscript{+2.7} \\
 + Alpaca 7b queries & \textbf{40.0\textsuperscript{+1.2}} & \textbf{89.0\textsuperscript{+1.2}} & \textbf{99.1} & \textbf{72.9\textsuperscript{+1.8}} & 71.3\textsuperscript{+2.9} \\
 + Alpaca 13b queries & 39.6\textsuperscript{+0.8} & 88.8\textsuperscript{+1.0} & 98.9 & 72.6\textsuperscript{+1.5} & \textbf{72.3\textsuperscript{+3.9}} \\
\midrule
\multicolumn{6}{l}{\textbf{\textit{Bottlenecked Query Generation}}} \\ 
\midrule
Baseline & 39.3 & 87.9 & 98.6 & 69.9 & 67.4 \\
 + tk-instruct 3b queries & \textbf{40.3\textsuperscript{+1.0}} & \textbf{88.7\textsuperscript{+0.8}} & \textbf{98.9} & 70.7\textsuperscript{+0.8} & 70.0\textsuperscript{+2.6} \\
 + Alpaca 7b queries & 39.9\textsuperscript{+0.6} & 88.2\textsuperscript & 98.7 & 69.6 & \textbf{70.7\textsuperscript{+3.3}} \\
 + Alpaca 13b queries & 39.7 & 88.3\textsuperscript & 98.7 & \textbf{70.8\textsuperscript{+0.9}} & 69.4\textsuperscript{+2.0} \\
\bottomrule
\end{tabular}
\caption{
Fine-tuned results of pre-training with LLM-based document expansion.
$\dagger$ denotes our reproduced results. The best scores are marked in \textbf{bold}. 
Results with the increment over the corresponding baseline have been tested with two-tailed t-tests, demonstrating statistically significant improvements ( p-value $\leq$ 0.01 ).
}
\label{table_results_main}
\end{table*}

\subsection{Zero-shot evaluation and Fine-tuning}
We conduct the zero-shot evaluation of the contrastive pre-trained encoder without fine-tuning on MS-MARCO, TREC-DL, and BEIR datasets. 
We conduct fine-tuning on both pre-training schemas to verify their retrieval initialization ability. Following DPR \cite{karpukhin2020dense}, a simple contrastive loss is applied to optimize the retriever.

\begin{equation}
\mathcal{L}=-\log \frac{\exp(q \cdot p^{+})}{\exp(q \cdot p^{+}) + \sum \exp(q \cdot p^{-})}
\end{equation}
where $q$ is a given query, $p^+$ and $p^-$ are their corresponding positive passage and negative passages respectively. 

\section{Experiments}
This section introduces detailed experiment settings for pre-training and fine-tuning. Then we present the main results.

\subsection{Pre-training}
Following the pretraining settings in \cite{gao2022unsupervised}, we choose the MS-MARCO dataset \cite{tri2016msmarco} with 3.2M documents as our pre-training corpus. LLMs with different types and parameter sizes, i.e. Alpaca 7B, 13B \cite{wang2023selfInst}, and tk-instruct 3B \cite{Yizhong2022SuperNaturalInstructions} are used to generate the queries for LLM-based document expansion.
Nucleus sampling with $top_p=0.95$, $top_k=50$, and $temperature=0.7$ is used for LLM generation.

For bottlenecked query generation pre-training, the encoder is initialized from the 12-layer BERT-\textit{base} model \cite{devlin2019bert}, while the single-layer decoder is randomly initialized from scratch. We use the AdamW optimizer with a learning rate of 3e-4, batch size of 2048, total steps of 80k, and a warmup ratio of 0.1. The pre-training uses 8 Tesla A100 GPUs and trains for 19 hours. For contrastive pre-training, we adapt the codes and architecture from \cite{gao2022unsupervised} and initialize from \cite{gao2021condenser} by following their settings. We use a learning rate of 1e-4, batch size of 2048, and total steps of 120k and keep other hyper-parameters the same as above for training 50 hours. For curriculum learning, 75\% of the total steps are used for the first stage of pre-training with sampled spans, and the remaining 25\% of the steps are used for the second stage of pre-training with LLM-generated queries. We use the cosine scheduler with the same hyper-parameter settings for the first stage, and a constant learning rate for the second stage. All pre-training seeds are set to 42 for reproducibility. The encoders are directly tested on downstream tasks without fine-tuning for zero-shot evaluation.

\begin{table*}[!ht]
    \centering
    \resizebox{\linewidth}{!}{
    \begin{tabular}{l|c|ccc|cccc}
    \toprule
        \textbf{Results / nDCG@10} & BM25 & coCondenser & Contriever & SimCSE & Baseline &  + tk-Instruct 3b &  + Alpaca 7b &  + Alpaca 13b \\ \midrule
        TREC-COVID & \textbf{65.6}  & 21.2  & 27.3  & 27.5  & 16.2  & 36.8\textsuperscript{+20.6}  & 52.3\textsuperscript{+36.1}  & 54.7\textsuperscript{+38.5}  \\ 
        NFCorpus & 32.5  & 13.7  & 31.7  & 10.5  & 29.9  & 33.1\textsuperscript{+3.2}  & 30.9\textsuperscript{+1.0}  & \textbf{33.5\textsuperscript{+3.5}}  \\ \midrule
        NQ & 32.9  & 10.7  & 25.4  & 16.3  & 9.3  & \textbf{34.3\textsuperscript{+25.0}}  & 31.8\textsuperscript{+22.5}  & 31.9\textsuperscript{+22.6}  \\ 
        HotpotQA & \textbf{60.3}  & 22.3  & 48.1  & 23.8  & 24.2  & 56.2\textsuperscript{+32.0}  & 51.5\textsuperscript{+27.3}  & 51.8\textsuperscript{+27.6}  \\ 
        FiQA-2018 & 23.6  & 7.2  & 24.5  & 9.7  & 19.6  & \textbf{29.8\textsuperscript{+10.3}}  & 27.2\textsuperscript{+7.6}  & 28.6\textsuperscript{+9.0}  \\ \midrule
        ArguAna & 31.5  & 34.4  & 37.9  & 28.0  & 35.8  & \textbf{44.6\textsuperscript{+8.8}}  & 40.5\textsuperscript{+4.8}  & 40.6\textsuperscript{+4.9}  \\ 
        Touché-2020 & \textbf{36.7}  & 5.8  & 16.7  & 13.4  & 8.1  & 16.3\textsuperscript{+8.2}  & 13.7\textsuperscript{+5.5}  & 16.9\textsuperscript{+8.7}  \\ \midrule
        CQADupStack & 29.9  & 10.5  & 28.4  & 13.5  & 18.2  & 30.9\textsuperscript{+12.8}  & 32.4\textsuperscript{+14.2}  & \textbf{33.3\textsuperscript{+15.1}}  \\ 
        Quora & 78.9  & 71.3  & 83.5  & 73.7  & 75.8  & 83.8\textsuperscript{+8.0}  & 83.3\textsuperscript{+7.5}  & \textbf{84.3\textsuperscript{+8.5}}  \\ \midrule
        DBPedia & \textbf{31.3}  & 16.3  & 29.2  & 16.7  & 22.5  & 30.2\textsuperscript{+7.7}  & 28.8\textsuperscript{+6.3}  & 29.6\textsuperscript{+7.1}  \\ \midrule
        SCIDOCS & \textbf{15.8}  & 4.6  & 14.9  & 6.1  & 10.4  & 13.6\textsuperscript{+3.2}  & 13.5\textsuperscript{+3.2}  & 14.4\textsuperscript{+4.1}  \\ \midrule
        FEVER & \textbf{75.3}  & 16.8  & 68.2  & 29.2  & 43.6  & 61.9\textsuperscript{+18.3}  & 67.2\textsuperscript{+23.6}  & 73.1\textsuperscript{+29.5}  \\ 
        Climate-FEVER & \textbf{21.3}  & 6.4  & 15.5  & 14.2  & 8.5  & 18.4\textsuperscript{+9.8}  & 13.8\textsuperscript{+5.3}  & 17.2\textsuperscript{+8.6}  \\ 
        SciFact & \textbf{66.5}  & 43.2  & 64.9  & 25.0  & 52.7  & 64.4\textsuperscript{+11.7}  & 60.8\textsuperscript{+8.1}  & 60.9\textsuperscript{+8.2}  \\ \midrule
        Average & \textbf{43.0}  & 20.3  & 36.9  & 22.0  & 26.8  & 39.6\textsuperscript{+12.8}  & 39.1\textsuperscript{+12.4}  & 40.8\textsuperscript{+14.0} \\ 
    \bottomrule
    \end{tabular}
    }
    \caption{
Out-of-domain zero-shot evaluation of contrastive pre-training with LLM-based document expansion on BEIR benchmark. 
All baselines tested on nDCG@10 are based on our reproduction.
Results with the increment over the corresponding baseline have been tested with two-tailed t-tests, demonstrating statistically significant improvements ( p-value $\leq$ 0.01 ).
}
\label{table_beir_zero_shot}
\end{table*}

\subsection{Fine-tuning}
The encoder is fine-tuned and tested on MS-MARCO Passage Ranking task \cite{tri2016msmarco}, TREC Deep Learning (DL) 2019 \cite{craswell2020trec19} and 2020 \cite{craswell2021trec20}.
MS-MARCO Passage Ranking dataset contains 8.8 million passages and 500k human annotated query-passage pairs. Following \cite{gao2021condenser}, we report the performance metrics on MRR@10, Recall@50, Recall@1K, and evaluate the models on its development set with 6,980 queries, because its test set is not publicly available. 
TREC-DL 2019 and 2020 test sets both contain 200 annotated queries.
We adopt the Tevatron pipeline \cite{gao2022tevatron} with the AdamW optimizer for a learning rate of 2e-5, a batch size of 8, negative samples per passage of 15, a negative depth of 200, and trains for 3 epochs. 
The performance metrics of TREC and BEIR are reported on NDCG@10. 

\subsection{Baselines}
We compare to self-contained baselines without using LLM-expanded queries, but only use randomly sampled spans as coarse-grained contexts. All other hyper-parameters used in the pre-training remain the same as the main experiments for fair comparison. 
In fine-tuned experiments, the contrastive pre-training baselines are mainly from \cite{wu2022query-as-context} by following their hyper-parameter settings, and other baselines are based on our settings. 

We also compare with other remarkable baselines, including the traditional sparse retrieval BM25 \cite{robertson2009probabilistic}, unsupervised sentence similarity encoder SimCSE \cite{gao2021simcse}, unsupervised contrastive pre-training method coCondenser \cite{gao2022unsupervised} and Contriever \cite{Izacard2021contriever} for zero-shot evaluation. For fine-tuned results, we also compare with the latest bottlenecked pre-training methods, including Condenser \cite{gao2021condenser}, SimLM \cite{wang2022simlm}, RetroMAE \cite{liu2022retromae} and CoT-MAE \cite{wu2023contextual}. Note that the recent bottlenecked methods using multi-task pre-training \cite{zhou2022master} or hybrid retrieval \cite{liu2023retromaev2, wu2023cotmaev2} are not compared, as they are beyond the scope of fair comparison.

\subsection{Zero-shot Evaluation}
Table \ref{table_results_zero_shot} reports the \textit{in-domain} zero-shot evaluation of contrastive pre-training with LLM-based document expansion. Pre-training with LLM-expanded queries shows clear improvements over its baselines that merely use randomly sampled passages. This indicates that our method achieves strong zero-shot retrieval abilities for in-domain evaluation on the MS-MARCO and TREC-DL 19 \& 20 datasets.

\subsection{Fine-tuned Retrieval}
The fine-tuned results of the two pre-training methods, i.e., contrastive pretraining and bottlenecked query generation pretraining, are presented in Table \ref{table_results_main}. Pre-training with LLM-expanded queries also gives a statistically significant boost to their baselines and counterparts. In addition, we notice that 1) Contrastive pre-training gives better results on the MS-MARCO passage task (in Recall@50 and Recall@1k) and TREC-DL 19 \& 20 (in nDCG@10). 2) Bottlenecked query generation gives better initialization on MS-MARCO w.r.t the official preferred metric MRR@10, but still lies behind contrastive pre-training in other metrics. 

\subsection{Out-of-domain Evaluation}
We also evaluate the out-of-domain zero-shot BEIR benchmark for contrastive pre-training with LLM-based document expansion and report the metric (nDCG@10) in Table \ref{table_beir_zero_shot}. BM25 is a very strong baseline w.r.t all the other contrastive pre-training methods that do not go through human-labeled fine-tuning. Nevertheless, our method still shows strong improvements over its contrastive baseline. Specifically, compared with Contriever \cite{Izacard2021contriever}, which is an unsupervised contrastive method pre-trained on a much larger corpus CCNET \cite{Wenzek2020ccnet}, pre-training with LLM expansion also shows superior retrieval performances. 


\section{Extended Analyses}
In this section, we analyze the effect of scaling up LLMs and the curriculum learning strategy with expanded queries generated by Alpaca 13b \footnote{Alpaca 13b is chosen because of better results in zero-shot and on-par performances in fine-tuned retrieval.}.

\begin{figure}[ht!]
\centering
\includegraphics[width=\linewidth]{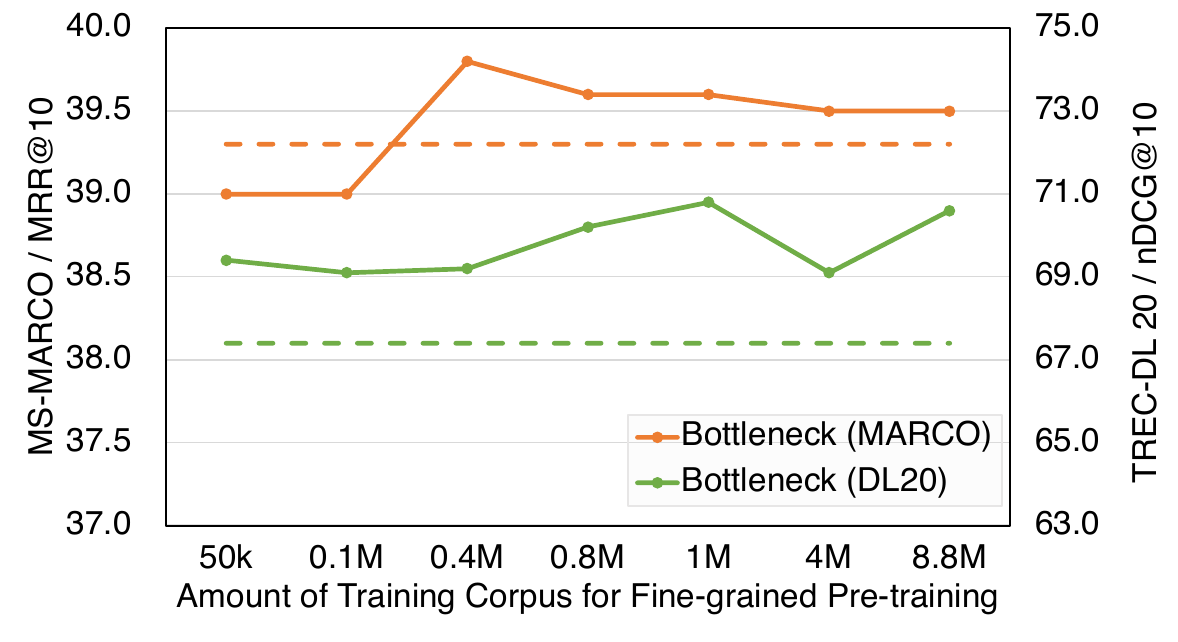}
\caption{
Effects of curriculum learning for fine-tuned bottlenecked pre-training with expanded queries generated by Alpaca 13b. The dashed lines are the corresponding baselines from Table \ref{table_results_main}. 
}
\label{llm_curriculum_bottleneck}
\end{figure}

\subsection{Effects of Scaling up LLMs}
We use three LLMs with different parameter sizes ranging from 3b to 13b, prompting them for document expansion and integrating the generated queries into pre-training. As shown in Table \ref{table_results_zero_shot}, scaling up the LLMs shows better retrieval performances in zero-shot contrastive pre-training. But this observation is not valid after fine-tuning in Table \ref{table_results_main}. We hypothesize that for fine-tuning with human labels, these LLMs are all capable enough for giving a good initialization for retrieval. 

\subsection{Effects of Curriculum Learning}
\begin{figure}[ht!]
\centering
\includegraphics[width=\linewidth]{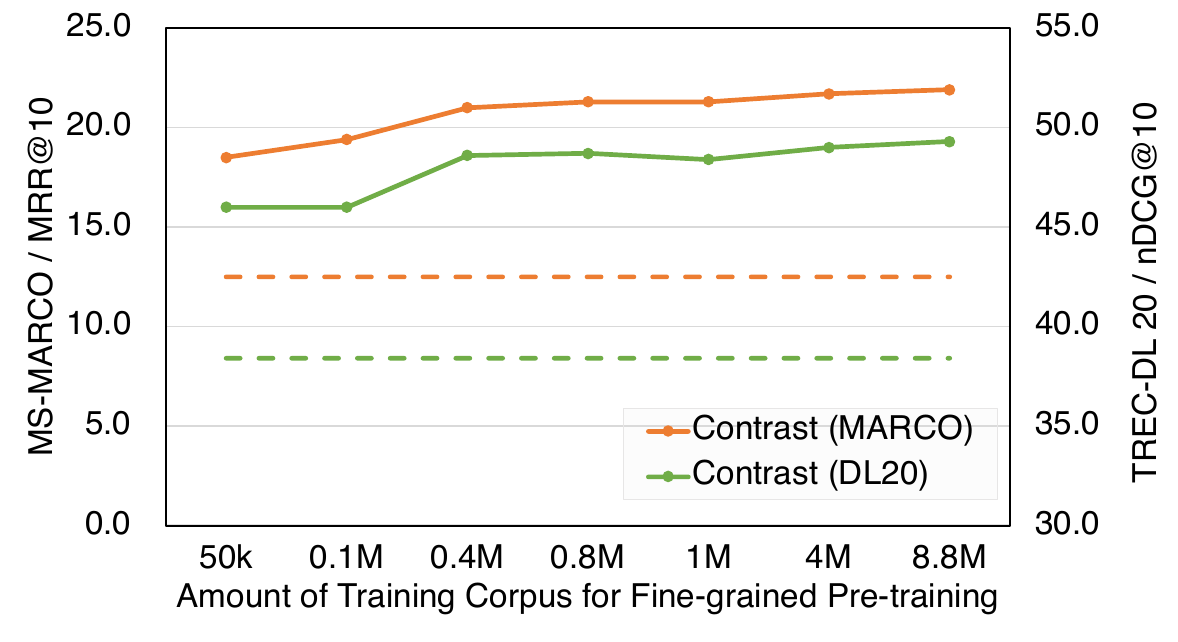}
\caption{
Effects of curriculum learning for zero-shot contrastive pre-training with LLM-expanded queries. 
}
\label{llm_curriculum_contrast}
\end{figure}

To further reduce the need for LLM-expanded queries in pre-training, we attempt to use a curriculum learning strategy as detailed before. We use randomly sampled spans as the coarse-grained context in the first stage of curriculum pre-training for 75\% of the total training steps. Then we use a small amount of LLM-expanded queries as the fine-grained context for the remaining pre-training steps. Figure \ref{llm_curriculum_bottleneck} and \ref{llm_curriculum_contrast} show that both pre-training schemas benefit from curriculum learning. 
Bottleneck query generation outperforms its baseline with just 0.4 million LLM-expanded queries after fine-tuning. Zero-shot contrastive pre-training surpasses the baselines and continues to demonstrate sustainable improvements as the number of fine-grained queries increases. 

\section{Related Works}
\subsection{Pre-training for Dense Retrieval}
Dense passage retrieval has gained sustainable improvements with the recent development of pre-training tasks. Some works focus on contrastive pre-training with constructed span relationship \cite{cheng2020pre}, randomly cropped spans \cite{gao2022unsupervised} or multiple granularity alignments \cite{ma2022pre}. And meanwhile, the others focus on pre-training with auxiliary bottlenecked decoders, like pre-training with a weak generative decoder \cite{lu2021less}, extreme masked ratio \cite{liu2022retromae}, and contextual span sampling \cite{wu2023contextual}. Our method is similar to \cite{gao2022unsupervised} and \cite{wu2023contextual}, but our core contribution is the methodology of incorporating expanded queries generated by LLMs into such pre-training schemas, which brings better context alignment and stronger zero-shot and fine-tuned performances.

\subsection{LLM-based Query and Document Expansion}
Traditional query or document expansions generate additional context via query rewriting \cite{lavrenko2017relevance_feedback}, or with specially fine-tuned T5 \cite{Rodrigo2019doc2query} or BART models \cite{Cho2022QueryGeneration}. With the bloom of LLMs \cite{long2022instructgpt, Hugo2023LLaMA, Yizhong2022SuperNaturalInstructions}, growing researches focus on using LLMs as query expansion models \cite{gao2023hyde, wang2023Query2doc, jagerman2023QueryExpension, yu2023Generate}, which enhance the lexical match of query-passage pairs. 

However, as discussed before, LLM-based document expansion is yet lacking exploration due to expensive inference costs brought by the huge amount of documents and the online inference issue. We propose to tackle those issues with pre-training techniques and curriculum learning strategies tailored for dense retrieval. Our method is also orthogonal to traditional query and document expansion and can incorporate them into the retrieval stage.

\section{Conclusion}
This paper systematically studies the potential of pre-training with Large Language Model-based document expansion for dense passage retrieval. Strong improvements in zero-shot and out-of-domain performances are observed in contrastive pre-training with LLM-based document expansion. Moreover, both contrastive pretraining and bottlenecked query generation pretraining achieve good retrieval abilities after fine-tuning. We further propose a two-stage curriculum learning strategy that can greatly reduce the need for LLM-expanded queries in pre-training, while keeping the minor performance degeneration. LLMs excel in expanding high-quality queries with enriched context information, which is suitable for scenarios lacking in human annotations. Researchers can thus deploy quick initialization of an unsupervised dense retrieval system with the pre-training of LLM-based document expansion, with even \textit{NO} human labels provided. 

\section{Limitation}
We are also interested in testing more types of LLMs with different sizes, such as ChatGPT \cite{long2022instructgpt}, and LLaMA 2 \cite{Hugo2023LLaMA}, or different prompts for document expansion, but our experiment budget is limited to support immediate investigations and we leave that to our future works.

\bibliography{aaai24}

\begin{thebibliography}{46}
\providecommand{\natexlab}[1]{#1}

\bibitem[{Bengio et~al.(2009)Bengio, Louradour, Collobert, and
  Weston}]{Bengio2009curriculumLearning}
Bengio, Y.; Louradour, J.; Collobert, R.; and Weston, J. 2009.
\newblock Curriculum learning.
\newblock In Danyluk, A.~P.; Bottou, L.; and Littman, M.~L., eds.,
  \emph{Proceedings of the 26th Annual International Conference on Machine
  Learning, {ICML} 2009, Montreal, Quebec, Canada, June 14-18, 2009}, volume
  382 of \emph{{ACM} International Conference Proceeding Series}, 41--48.
  {ACM}.

\bibitem[{Cai et~al.(2022)Cai, Wang, Liu, and Shi}]{cai2022recent}
Cai, D.; Wang, Y.; Liu, L.; and Shi, S. 2022.
\newblock Recent advances in retrieval-augmented text generation.
\newblock In \emph{Proceedings of the 45th International ACM SIGIR Conference
  on Research and Development in Information Retrieval}, 3417--3419.

\bibitem[{Chang et~al.(2020)Chang, Yu, Chang, Yang, and Kumar}]{cheng2020pre}
Chang, W.; Yu, F.~X.; Chang, Y.; Yang, Y.; and Kumar, S. 2020.
\newblock Pre-training Tasks for Embedding-based Large-scale Retrieval.
\newblock In \emph{8th International Conference on Learning Representations,
  {ICLR} 2020, Addis Ababa, Ethiopia, April 26-30, 2020}. OpenReview.net.

\bibitem[{Cho et~al.(2022)Cho, Jeong, Yang, and Park}]{Cho2022QueryGeneration}
Cho, S.; Jeong, S.; Yang, W.; and Park, J.~C. 2022.
\newblock Query Generation with External Knowledge for Dense Retrieval.
\newblock In Agirre, E.; Apidianaki, M.; and Vulic, I., eds., \emph{Proceedings
  of Deep Learning Inside Out: The 3rd Workshop on Knowledge Extraction and
  Integration for Deep Learning Architectures, DeeLIO@ACL 2022, Dublin, Ireland
  and Online, May 27, 2022}, 22--32. Association for Computational Linguistics.

\bibitem[{Chowdhery et~al.(2022)Chowdhery, Narang, Devlin, Bosma, Mishra,
  Roberts, Barham, Chung, Sutton, Gehrmann, Schuh, Shi, Tsvyashchenko, Maynez,
  Rao, Barnes, Tay, Shazeer, Prabhakaran, Reif, Du, Hutchinson, Pope, Bradbury,
  Austin, Isard, Gur{-}Ari, Yin, Duke, Levskaya, Ghemawat, Dev, Michalewski,
  Garcia, Misra, Robinson, Fedus, Zhou, Ippolito, Luan, Lim, Zoph, Spiridonov,
  Sepassi, Dohan, Agrawal, Omernick, Dai, Pillai, Pellat, Lewkowycz, Moreira,
  Child, Polozov, Lee, Zhou, Wang, Saeta, Diaz, Firat, Catasta, Wei,
  Meier{-}Hellstern, Eck, Dean, Petrov, and Fiedel}]{Aakanksha2022PaLM}
Chowdhery, A.; Narang, S.; Devlin, J.; Bosma, M.; Mishra, G.; Roberts, A.;
  Barham, P.; Chung, H.~W.; Sutton, C.; Gehrmann, S.; Schuh, P.; Shi, K.;
  Tsvyashchenko, S.; Maynez, J.; Rao, A.; Barnes, P.; Tay, Y.; Shazeer, N.;
  Prabhakaran, V.; Reif, E.; Du, N.; Hutchinson, B.; Pope, R.; Bradbury, J.;
  Austin, J.; Isard, M.; Gur{-}Ari, G.; Yin, P.; Duke, T.; Levskaya, A.;
  Ghemawat, S.; Dev, S.; Michalewski, H.; Garcia, X.; Misra, V.; Robinson, K.;
  Fedus, L.; Zhou, D.; Ippolito, D.; Luan, D.; Lim, H.; Zoph, B.; Spiridonov,
  A.; Sepassi, R.; Dohan, D.; Agrawal, S.; Omernick, M.; Dai, A.~M.; Pillai,
  T.~S.; Pellat, M.; Lewkowycz, A.; Moreira, E.; Child, R.; Polozov, O.; Lee,
  K.; Zhou, Z.; Wang, X.; Saeta, B.; Diaz, M.; Firat, O.; Catasta, M.; Wei, J.;
  Meier{-}Hellstern, K.; Eck, D.; Dean, J.; Petrov, S.; and Fiedel, N. 2022.
\newblock PaLM: Scaling Language Modeling with Pathways.
\newblock \emph{CoRR}, abs/2204.02311.

\bibitem[{Craswell et~al.(2021)Craswell, Mitra, Yilmaz, and
  Campos}]{craswell2021trec20}
Craswell, N.; Mitra, B.; Yilmaz, E.; and Campos, D. 2021.
\newblock Overview of the TREC 2020 deep learning track.
\newblock arXiv:2102.07662.

\bibitem[{Craswell et~al.(2020)Craswell, Mitra, Yilmaz, Campos, and
  Voorhees}]{craswell2020trec19}
Craswell, N.; Mitra, B.; Yilmaz, E.; Campos, D.; and Voorhees, E.~M. 2020.
\newblock Overview of the TREC 2019 deep learning track.
\newblock arXiv:2003.07820.

\bibitem[{Devlin et~al.(2019)Devlin, Chang, Lee, and
  Toutanova}]{devlin2019bert}
Devlin, J.; Chang, M.-W.; Lee, K.; and Toutanova, K. 2019.
\newblock {BERT}: Pre-training of Deep Bidirectional Transformers for Language
  Understanding.
\newblock In \emph{Proceedings of the 2019 Conference of the North {A}merican
  Chapter of the Association for Computational Linguistics: Human Language
  Technologies, Volume 1 (Long and Short Papers)}, 4171--4186. Minneapolis,
  Minnesota: Association for Computational Linguistics.

\bibitem[{Gao and Callan(2021)}]{gao2021condenser}
Gao, L.; and Callan, J. 2021.
\newblock Condenser: a Pre-training Architecture for Dense Retrieval.
\newblock In \emph{Proceedings of the 2021 Conference on Empirical Methods in
  Natural Language Processing}, 981--993. Online and Punta Cana, Dominican
  Republic: Association for Computational Linguistics.

\bibitem[{Gao and Callan(2022)}]{gao2022unsupervised}
Gao, L.; and Callan, J. 2022.
\newblock Unsupervised Corpus Aware Language Model Pre-training for Dense
  Passage Retrieval.
\newblock In \emph{Proceedings of the 60th Annual Meeting of the Association
  for Computational Linguistics (Volume 1: Long Papers)}, 2843--2853. Dublin,
  Ireland: Association for Computational Linguistics.

\bibitem[{Gao et~al.(2022)Gao, Ma, Lin, and Callan}]{gao2022tevatron}
Gao, L.; Ma, X.; Lin, J.; and Callan, J. 2022.
\newblock Tevatron: An efficient and flexible toolkit for dense retrieval.
\newblock \emph{arXiv preprint arXiv:2203.05765}.

\bibitem[{Gao et~al.(2023)Gao, Ma, Lin, and Callan}]{gao2023hyde}
Gao, L.; Ma, X.; Lin, J.; and Callan, J. 2023.
\newblock Precise Zero-Shot Dense Retrieval without Relevance Labels.
\newblock In Rogers, A.; Boyd{-}Graber, J.~L.; and Okazaki, N., eds.,
  \emph{Proceedings of the 61st Annual Meeting of the Association for
  Computational Linguistics (Volume 1: Long Papers), {ACL} 2023, Toronto,
  Canada, July 9-14, 2023}, 1762--1777. Association for Computational
  Linguistics.

\bibitem[{Gao, Yao, and Chen(2021)}]{gao2021simcse}
Gao, T.; Yao, X.; and Chen, D. 2021.
\newblock {S}im{CSE}: Simple Contrastive Learning of Sentence Embeddings.
\newblock In \emph{Proceedings of the 2021 Conference on Empirical Methods in
  Natural Language Processing}, 6894--6910. Online and Punta Cana, Dominican
  Republic: Association for Computational Linguistics.

\bibitem[{Izacard et~al.(2021)Izacard, Caron, Hosseini, Riedel, Bojanowski,
  Joulin, and Grave}]{Izacard2021contriever}
Izacard, G.; Caron, M.; Hosseini, L.; Riedel, S.; Bojanowski, P.; Joulin, A.;
  and Grave, E. 2021.
\newblock Towards Unsupervised Dense Information Retrieval with Contrastive
  Learning.
\newblock \emph{CoRR}, abs/2112.09118.

\bibitem[{Jagerman et~al.(2023)Jagerman, Zhuang, Qin, Wang, and
  Bendersky}]{jagerman2023QueryExpension}
Jagerman, R.; Zhuang, H.; Qin, Z.; Wang, X.; and Bendersky, M. 2023.
\newblock Query Expansion by Prompting Large Language Models.
\newblock \emph{CoRR}, abs/2305.03653.

\bibitem[{Karpukhin et~al.(2020)Karpukhin, Oguz, Min, Lewis, Wu, Edunov, Chen,
  and Yih}]{karpukhin2020dense}
Karpukhin, V.; Oguz, B.; Min, S.; Lewis, P.; Wu, L.; Edunov, S.; Chen, D.; and
  Yih, W.-t. 2020.
\newblock Dense Passage Retrieval for Open-Domain Question Answering.
\newblock In \emph{Proceedings of the 2020 Conference on Empirical Methods in
  Natural Language Processing (EMNLP)}, 6769--6781. Online: Association for
  Computational Linguistics.

\bibitem[{Khattab and Zaharia(2020)}]{Omar2020colbert}
Khattab, O.; and Zaharia, M. 2020.
\newblock ColBERT: Efficient and Effective Passage Search via Contextualized
  Late Interaction over {BERT}.
\newblock In Huang, J.~X.; Chang, Y.; Cheng, X.; Kamps, J.; Murdock, V.; Wen,
  J.; and Liu, Y., eds., \emph{Proceedings of the 43rd International {ACM}
  {SIGIR} conference on research and development in Information Retrieval,
  {SIGIR} 2020, Virtual Event, China, July 25-30, 2020}, 39--48. {ACM}.

\bibitem[{Lavrenko and Croft(2017)}]{lavrenko2017relevance_feedback}
Lavrenko, V.; and Croft, W.~B. 2017.
\newblock Relevance-Based Language Models.
\newblock \emph{{SIGIR} Forum}, 51(2): 260--267.

\bibitem[{Lewis et~al.(2020)Lewis, Perez, Piktus, Petroni, Karpukhin, Goyal,
  K{\"{u}}ttler, Lewis, Yih, Rockt{\"{a}}schel, Riedel, and
  Kiela}]{Lewis2020Retrieval}
Lewis, P. S.~H.; Perez, E.; Piktus, A.; Petroni, F.; Karpukhin, V.; Goyal, N.;
  K{\"{u}}ttler, H.; Lewis, M.; Yih, W.; Rockt{\"{a}}schel, T.; Riedel, S.; and
  Kiela, D. 2020.
\newblock Retrieval-Augmented Generation for Knowledge-Intensive {NLP} Tasks.
\newblock In Larochelle, H.; Ranzato, M.; Hadsell, R.; Balcan, M.; and Lin, H.,
  eds., \emph{Advances in Neural Information Processing Systems 33: Annual
  Conference on Neural Information Processing Systems 2020, NeurIPS 2020,
  December 6-12, 2020, virtual}.

\bibitem[{Liu et~al.(2021)Liu, Lu, Cheng, Shi, Wang, Cheng, and
  Yin}]{Liu2021baidu_search}
Liu, Y.; Lu, W.; Cheng, S.; Shi, D.; Wang, S.; Cheng, Z.; and Yin, D. 2021.
\newblock Pre-trained Language Model for Web-scale Retrieval in Baidu Search.
\newblock In Zhu, F.; Ooi, B.~C.; and Miao, C., eds., \emph{{KDD} '21: The 27th
  {ACM} {SIGKDD} Conference on Knowledge Discovery and Data Mining, Virtual
  Event, Singapore, August 14-18, 2021}, 3365--3375. {ACM}.

\bibitem[{Liu and Shao(2022)}]{liu2022retromae}
Liu, Z.; and Shao, Y. 2022.
\newblock RetroMAE: Pre-training Retrieval-oriented Transformers via Masked
  Auto-Encoder.
\newblock \emph{arXiv preprint arXiv:2205.12035}.

\bibitem[{Liu et~al.(2023)Liu, Xiao, Shao, and Cao}]{liu2023retromaev2}
Liu, Z.; Xiao, S.; Shao, Y.; and Cao, Z. 2023.
\newblock RetroMAE-2: Duplex Masked Auto-Encoder For Pre-Training
  Retrieval-Oriented Language Models.
\newblock In Rogers, A.; Boyd{-}Graber, J.~L.; and Okazaki, N., eds.,
  \emph{Proceedings of the 61st Annual Meeting of the Association for
  Computational Linguistics (Volume 1: Long Papers), {ACL} 2023, Toronto,
  Canada, July 9-14, 2023}, 2635--2648. Association for Computational
  Linguistics.

\bibitem[{Lu et~al.(2021)Lu, He, Xiong, Ke, Malik, Dou, Bennett, Liu, and
  Overwijk}]{lu2021less}
Lu, S.; He, D.; Xiong, C.; Ke, G.; Malik, W.; Dou, Z.; Bennett, P.; Liu, T.-Y.;
  and Overwijk, A. 2021.
\newblock Less is More: Pretrain a Strong Siamese Encoder for Dense Text
  Retrieval Using a Weak Decoder.
\newblock In \emph{Proceedings of the 2021 Conference on Empirical Methods in
  Natural Language Processing}, 2780--2791.

\bibitem[{Lu et~al.(2022)Lu, Liu, Liu, Shi, Huang, Sun, Tian, Wu, Wang, Yin
  et~al.}]{lu2022ernie_search}
Lu, Y.; Liu, Y.; Liu, J.; Shi, Y.; Huang, Z.; Sun, S. F.~Y.; Tian, H.; Wu, H.;
  Wang, S.; Yin, D.; et~al. 2022.
\newblock Ernie-search: Bridging cross-encoder with dual-encoder via self
  on-the-fly distillation for dense passage retrieval.
\newblock \emph{arXiv preprint arXiv:2205.09153}.

\bibitem[{Ma et~al.(2022)Ma, Guo, Zhang, Fan, and Cheng}]{ma2022pre}
Ma, X.; Guo, J.; Zhang, R.; Fan, Y.; and Cheng, X. 2022.
\newblock Pre-train a Discriminative Text Encoder for Dense Retrieval via
  Contrastive Span Prediction.
\newblock \emph{arXiv preprint arXiv:2204.10641}.

\bibitem[{Nguyen et~al.(2016)Nguyen, Rosenberg, Song, Gao, Tiwary, Majumder,
  and Deng}]{tri2016msmarco}
Nguyen, T.; Rosenberg, M.; Song, X.; Gao, J.; Tiwary, S.; Majumder, R.; and
  Deng, L. 2016.
\newblock {MS} {MARCO:} {A} Human Generated MAchine Reading COmprehension
  Dataset.
\newblock In Besold, T.~R.; Bordes, A.; d'Avila Garcez, A.~S.; and Wayne, G.,
  eds., \emph{Proceedings of the Workshop on Cognitive Computation: Integrating
  neural and symbolic approaches 2016 co-located with the 30th Annual
  Conference on Neural Information Processing Systems {(NIPS} 2016), Barcelona,
  Spain, December 9, 2016}, volume 1773 of \emph{{CEUR} Workshop Proceedings}.
  CEUR-WS.org.

\bibitem[{Nogueira et~al.(2019)Nogueira, Yang, Lin, and
  Cho}]{Rodrigo2019doc2query}
Nogueira, R.~F.; Yang, W.; Lin, J.; and Cho, K. 2019.
\newblock Document Expansion by Query Prediction.
\newblock \emph{CoRR}, abs/1904.08375.

\bibitem[{Ouyang et~al.(2022)Ouyang, Wu, Jiang, Almeida, Wainwright, Mishkin,
  Zhang, Agarwal, Slama, Ray, Schulman, Hilton, Kelton, Miller, Simens, Askell,
  Welinder, Christiano, Leike, and Lowe}]{long2022instructgpt}
Ouyang, L.; Wu, J.; Jiang, X.; Almeida, D.; Wainwright, C.~L.; Mishkin, P.;
  Zhang, C.; Agarwal, S.; Slama, K.; Ray, A.; Schulman, J.; Hilton, J.; Kelton,
  F.; Miller, L.; Simens, M.; Askell, A.; Welinder, P.; Christiano, P.~F.;
  Leike, J.; and Lowe, R. 2022.
\newblock Training language models to follow instructions with human feedback.
\newblock In \emph{NeurIPS}.

\bibitem[{Qu et~al.(2021)Qu, Ding, Liu, Liu, Ren, Zhao, Dong, Wu, and
  Wang}]{qu-etal-2021-rocketqa}
Qu, Y.; Ding, Y.; Liu, J.; Liu, K.; Ren, R.; Zhao, W.~X.; Dong, D.; Wu, H.; and
  Wang, H. 2021.
\newblock {R}ocket{QA}: An Optimized Training Approach to Dense Passage
  Retrieval for Open-Domain Question Answering.
\newblock In \emph{Proceedings of the 2021 Conference of the North American
  Chapter of the Association for Computational Linguistics: Human Language
  Technologies}, 5835--5847. Online: Association for Computational Linguistics.

\bibitem[{Ren et~al.(2021)Ren, Qu, Liu, Zhao, She, Wu, Wang, and
  Wen}]{ren-etal-2021-rocketqav2}
Ren, R.; Qu, Y.; Liu, J.; Zhao, W.~X.; She, Q.; Wu, H.; Wang, H.; and Wen,
  J.-R. 2021.
\newblock {R}ocket{QA}v2: A Joint Training Method for Dense Passage Retrieval
  and Passage Re-ranking.
\newblock In \emph{Proceedings of the 2021 Conference on Empirical Methods in
  Natural Language Processing}, 2825--2835. Online and Punta Cana, Dominican
  Republic: Association for Computational Linguistics.

\bibitem[{Robertson, Zaragoza et~al.(2009)}]{robertson2009probabilistic}
Robertson, S.; Zaragoza, H.; et~al. 2009.
\newblock The probabilistic relevance framework: BM25 and beyond.
\newblock \emph{Foundations and Trends{\textregistered} in Information
  Retrieval}, 3(4): 333--389.

\bibitem[{Sakata et~al.(2019)Sakata, Shibata, Tanaka, and
  Kurohashi}]{Sakata2019FAQ}
Sakata, W.; Shibata, T.; Tanaka, R.; and Kurohashi, S. 2019.
\newblock {FAQ} Retrieval using Query-Question Similarity and BERT-Based
  Query-Answer Relevance.
\newblock In Piwowarski, B.; Chevalier, M.; Gaussier, {\'{E}}.; Maarek, Y.;
  Nie, J.; and Scholer, F., eds., \emph{Proceedings of the 42nd International
  {ACM} {SIGIR} Conference on Research and Development in Information
  Retrieval, {SIGIR} 2019, Paris, France, July 21-25, 2019}, 1113--1116. {ACM}.

\bibitem[{Santhanam et~al.(2022)Santhanam, Khattab, Saad-Falcon, Potts, and
  Zaharia}]{santhanam2022colbertv2}
Santhanam, K.; Khattab, O.; Saad-Falcon, J.; Potts, C.; and Zaharia, M. 2022.
\newblock {C}ol{BERT}v2: Effective and Efficient Retrieval via Lightweight Late
  Interaction.
\newblock In \emph{Proceedings of the 2022 Conference of the North American
  Chapter of the Association for Computational Linguistics: Human Language
  Technologies}, 3715--3734. Seattle, United States: Association for
  Computational Linguistics.

\bibitem[{Thakur et~al.(2021)Thakur, Reimers, R{\"{u}}ckl{\'{e}}, Srivastava,
  and Gurevych}]{thakur2021beir}
Thakur, N.; Reimers, N.; R{\"{u}}ckl{\'{e}}, A.; Srivastava, A.; and Gurevych,
  I. 2021.
\newblock {BEIR:} {A} Heterogenous Benchmark for Zero-shot Evaluation of
  Information Retrieval Models.
\newblock \emph{CoRR}, abs/2104.08663.

\bibitem[{Touvron et~al.(2023)Touvron, Lavril, Izacard, Martinet, Lachaux,
  Lacroix, Rozi{\`{e}}re, Goyal, Hambro, Azhar, Rodriguez, Joulin, Grave, and
  Lample}]{Hugo2023LLaMA}
Touvron, H.; Lavril, T.; Izacard, G.; Martinet, X.; Lachaux, M.; Lacroix, T.;
  Rozi{\`{e}}re, B.; Goyal, N.; Hambro, E.; Azhar, F.; Rodriguez, A.; Joulin,
  A.; Grave, E.; and Lample, G. 2023.
\newblock LLaMA: Open and Efficient Foundation Language Models.
\newblock \emph{CoRR}, abs/2302.13971.

\bibitem[{Wang et~al.(2022{\natexlab{a}})Wang, Yang, Huang, Jiao, Yang, Jiang,
  Majumder, and Wei}]{wang2022simlm}
Wang, L.; Yang, N.; Huang, X.; Jiao, B.; Yang, L.; Jiang, D.; Majumder, R.; and
  Wei, F. 2022{\natexlab{a}}.
\newblock SimLM: Pre-training with Representation Bottleneck for Dense Passage
  Retrieval.
\newblock \emph{CoRR}, abs/2207.02578.

\bibitem[{Wang, Yang, and Wei(2023)}]{wang2023Query2doc}
Wang, L.; Yang, N.; and Wei, F. 2023.
\newblock Query2doc: Query Expansion with Large Language Models.
\newblock \emph{CoRR}, abs/2303.07678.

\bibitem[{Wang et~al.(2023)Wang, Kordi, Mishra, Liu, Smith, Khashabi, and
  Hajishirzi}]{wang2023selfInst}
Wang, Y.; Kordi, Y.; Mishra, S.; Liu, A.; Smith, N.~A.; Khashabi, D.; and
  Hajishirzi, H. 2023.
\newblock Self-Instruct: Aligning Language Models with Self-Generated
  Instructions.
\newblock In Rogers, A.; Boyd{-}Graber, J.~L.; and Okazaki, N., eds.,
  \emph{Proceedings of the 61st Annual Meeting of the Association for
  Computational Linguistics (Volume 1: Long Papers), {ACL} 2023, Toronto,
  Canada, July 9-14, 2023}, 13484--13508. Association for Computational
  Linguistics.

\bibitem[{Wang et~al.(2022{\natexlab{b}})Wang, Mishra, Alipoormolabashi, Kordi,
  Mirzaei, Naik, Ashok, Dhanasekaran, Arunkumar, Stap, Pathak, Karamanolakis,
  Lai, Purohit, Mondal, Anderson, Kuznia, Doshi, Pal, Patel, Moradshahi,
  Parmar, Purohit, Varshney, Kaza, Verma, Puri, Karia, Doshi, Sampat, Mishra,
  A, Patro, Dixit, and Shen}]{Yizhong2022SuperNaturalInstructions}
Wang, Y.; Mishra, S.; Alipoormolabashi, P.; Kordi, Y.; Mirzaei, A.; Naik, A.;
  Ashok, A.; Dhanasekaran, A.~S.; Arunkumar, A.; Stap, D.; Pathak, E.;
  Karamanolakis, G.; Lai, H.~G.; Purohit, I.; Mondal, I.; Anderson, J.; Kuznia,
  K.; Doshi, K.; Pal, K.~K.; Patel, M.; Moradshahi, M.; Parmar, M.; Purohit,
  M.; Varshney, N.; Kaza, P.~R.; Verma, P.; Puri, R.~S.; Karia, R.; Doshi, S.;
  Sampat, S.~K.; Mishra, S.; A, S.~R.; Patro, S.; Dixit, T.; and Shen, X.
  2022{\natexlab{b}}.
\newblock Super-NaturalInstructions: Generalization via Declarative
  Instructions on 1600+ {NLP} Tasks.
\newblock In Goldberg, Y.; Kozareva, Z.; and Zhang, Y., eds., \emph{Proceedings
  of the 2022 Conference on Empirical Methods in Natural Language Processing,
  {EMNLP} 2022, Abu Dhabi, United Arab Emirates, December 7-11, 2022},
  5085--5109. Association for Computational Linguistics.

\bibitem[{Wenzek et~al.(2020)Wenzek, Lachaux, Conneau, Chaudhary, Guzm{\'{a}}n,
  Joulin, and Grave}]{Wenzek2020ccnet}
Wenzek, G.; Lachaux, M.; Conneau, A.; Chaudhary, V.; Guzm{\'{a}}n, F.; Joulin,
  A.; and Grave, E. 2020.
\newblock CCNet: Extracting High Quality Monolingual Datasets from Web Crawl
  Data.
\newblock In Calzolari, N.; B{\'{e}}chet, F.; Blache, P.; Choukri, K.; Cieri,
  C.; Declerck, T.; Goggi, S.; Isahara, H.; Maegaard, B.; Mariani, J.; Mazo,
  H.; Moreno, A.; Odijk, J.; and Piperidis, S., eds., \emph{Proceedings of The
  12th Language Resources and Evaluation Conference, {LREC} 2020, Marseille,
  France, May 11-16, 2020}, 4003--4012. European Language Resources
  Association.

\bibitem[{Wu, Ma, and Hu(2022)}]{wu2022query-as-context}
Wu, X.; Ma, G.; and Hu, S. 2022.
\newblock Query-as-context Pre-training for Dense Passage Retrieval.
\newblock \emph{CoRR}, abs/2212.09598.

\bibitem[{Wu et~al.(2023{\natexlab{a}})Wu, Ma, Lin, Lin, Wang, and
  Hu}]{wu2023contextual}
Wu, X.; Ma, G.; Lin, M.; Lin, Z.; Wang, Z.; and Hu, S. 2023{\natexlab{a}}.
\newblock ConTextual Masked Auto-Encoder for Dense Passage Retrieval.
\newblock In Williams, B.; Chen, Y.; and Neville, J., eds.,
  \emph{Thirty-Seventh {AAAI} Conference on Artificial Intelligence, {AAAI}
  2023, Thirty-Fifth Conference on Innovative Applications of Artificial
  Intelligence, {IAAI} 2023, Thirteenth Symposium on Educational Advances in
  Artificial Intelligence, {EAAI} 2023, Washington, DC, USA, February 7-14,
  2023}, 4738--4746. {AAAI} Press.

\bibitem[{Wu et~al.(2023{\natexlab{b}})Wu, Ma, Wang, Lin, Lin, Zhang, and
  Hu}]{wu2023cotmaev2}
Wu, X.; Ma, G.; Wang, P.; Lin, M.; Lin, Z.; Zhang, F.; and Hu, S.
  2023{\natexlab{b}}.
\newblock CoT-MAE v2: Contextual Masked Auto-Encoder with Multi-view Modeling
  for Passage Retrieval.
\newblock arXiv:2304.03158.

\bibitem[{Yu et~al.(2023)Yu, Iter, Wang, Xu, Ju, Sanyal, Zhu, Zeng, and
  Jiang}]{yu2023Generate}
Yu, W.; Iter, D.; Wang, S.; Xu, Y.; Ju, M.; Sanyal, S.; Zhu, C.; Zeng, M.; and
  Jiang, M. 2023.
\newblock Generate rather than Retrieve: Large Language Models are Strong
  Context Generators.
\newblock In \emph{The Eleventh International Conference on Learning
  Representations, {ICLR} 2023, Kigali, Rwanda, May 1-5, 2023}. OpenReview.net.

\bibitem[{Zhou et~al.(2022)Zhou, Liu, Gong, Zhao, Jiang, Duan, and
  Wen}]{zhou2022master}
Zhou, K.; Liu, X.; Gong, Y.; Zhao, W.~X.; Jiang, D.; Duan, N.; and Wen, J.-R.
  2022.
\newblock MASTER: Multi-task Pre-trained Bottlenecked Masked Autoencoders are
  Better Dense Retrievers.
\newblock \emph{arXiv preprint arXiv:2212.07841}.

\bibitem[{Zou et~al.(2023)Zou, Lu, Liu, Cai, Chu, Ma, Shi, Sun, Cheng, Gu,
  Wang, and Yin}]{Zou2023pre_trained}
Zou, L.; Lu, W.; Liu, Y.; Cai, H.; Chu, X.; Ma, D.; Shi, D.; Sun, Y.; Cheng,
  Z.; Gu, S.; Wang, S.; and Yin, D. 2023.
\newblock Pre-trained Language Model-based Retrieval and Ranking for Web
  Search.
\newblock \emph{{ACM} Trans. Web}, 17(1): 4:1--4:36.

\end{thebibliography}

\end{document}